\documentclass[conference]{IEEEtran}
\IEEEoverridecommandlockouts
\usepackage{cite}
\usepackage{amsmath,amssymb,amsfonts}
\usepackage{algorithmic}
\usepackage{graphicx}
\usepackage{textcomp}
\usepackage{xcolor}
\usepackage{booktabs}
\usepackage{graphicx}
\def\BibTeX{{\rm B\kern-.05em{\sc i\kern-.025em b}\kern-.08em
    T\kern-.1667em\lower.7ex\hbox{E}\kern-.125emX}}
\begin{document}

\title{OrthoDoc: Multimodal Large Language Model for Assisting Diagnosis in Computed Tomography}

\author{
\IEEEauthorblockN{Youzhu Jin}
	\IEEEauthorblockA{\textit{Beijing-Dublin International College} \\
		\textit{Beijing University of Technology}\\
		Beijing, China\\
		youzhu.jin@ucdconnect.ie}
\and
\IEEEauthorblockN{Yichen Zhang}
\IEEEauthorblockA{\textit{Beijing-Dublin International College} \\
	\textit{Beijing University of Technology}\\
	Beijing, China\\
	yichen.zhang2@ucdconnect.ie}
}
\maketitle

\begin{abstract}
Multimodal large language models (MLLMs) have achieved significant success in the general field of image processing. Their emerging task generalization and free-form conversational capabilities can greatly facilitate medical diagnostic assistance, helping patients better understand their conditions and enhancing doctor-patient trust. Computed Tomography (CT) is a non-invasive imaging technique used to capture the internal mechanisms of a patient's condition and is widely utilized. However, in past research, the complex textural features of this imaging data have made accurate interpretation by algorithms challenging, impeding the performance of general LLMs in diagnostic assistance. To address this, we developed OrthoDoc, a MLLM designed for CT diagnostics. OrthoDoc is trained on 120,000 CT images and diagnostic reports and includes a Retrieval-Augmented Generation(RAG) module capable of effectively mitigating model hallucinations. This module is informed by extensive medical literature, textbooks, and explanatory data. Thus, OrthoDoc not only processes complex CT images but also stores, understands, and reasons over medical knowledge and language. In extensive experiments, OrthoDoc outperforms commercial models led by GPT-4, demonstrating superior diagnostic capabilities and accuracy. Specifically, OrthoDoc significantly surpasses existing models in the diagnosis of common orthopedic conditions such as fractures, arthritis, and tumors. Additionally, OrthoDoc exhibits robust generalization and stability when handling rare and complex cases.
\end{abstract}


\section{Introduction}

With the rapid advancement of artificial intelligence technology, there is a growing research interest in using AI systems to assist in medical image diagnostics. \cite{h1,h2,h3,h4} Ideally, AI models can handle multiple modalities of medical data, including patients' basic vital signs, pathological slide data, and computed tomography (CT) data. These systems aim to assist doctors in real-world diagnostics by engaging in natural language dialogue, thus providing comprehensive support across various medical scenarios.

Traditional AI models for medical image interpretation often fall short of the precision required for effective diagnostic assistance. These models are typically limited to specific tasks such as medical image classification and segmentation, which restricts their utility in broader diagnostic contexts.\cite{h5,h6,h7} Moreover, they often lack the capability to engage in free-form conversational interactions, which is crucial for nuanced medical consultations. Multimodal large language models (MLLMs), exemplified by ChatGPT, possess powerful natural language understanding and generation capabilities. Their ability to handle multimodal data—integrating text, images, and other forms of data—opens up new possibilities for advanced diagnostic assistance.

However, the intelligent interpretation of CT images presents unique challenges. Unlike general image recognition tasks, medical image analysis requires models to have extensive prior knowledge of complex medical images. Specifically, fractures exhibit distinct features on CT images, and models must not only accurately identify these features but also discern subtle differences among various types of fractures and understand their clinical significance. Traditional models often lack this depth of medical knowledge, leading to suboptimal performance in real-world applications.

Additionally, the textual information used in CT medical diagnostics is highly specialized and complex. In a series of zero-shot experiments, representative open-source and commercial models frequently encountered significant hallucination issues. These models often generated inaccurate or misleading content when dealing with medical domain-specific terminology and complex diagnostic reports. Such inaccuracies pose significant risks to medical decision-making, highlighting the need for models that can generate precise and reliable outputs.

To address these challenges, we developed OrthoDoc, a multimodal large model specifically designed for CT diagnostics. OrthoDoc is trained on a diverse dataset comprising 120,000 CT images and their corresponding diagnostic reports. This extensive training enables the model to master a wide range of medical imaging features and acquire comprehensive diagnostic knowledge. We incorporated medical professional texts from a broad array of data sources, particularly those rich in orthopedic and diagnostic knowledge. Additionally, we integrated an automated knowledge graph construction and retrieval component. This enhancement allows the model to support its text reasoning process with a robust RAG module, effectively eliminating severe hallucinations.

OrthoDoc is capable of processing complex CT images and generating detailed diagnostic reports in natural language, offering valuable diagnostic information and treatment recommendations for physicians. This multimodal capability allows OrthoDoc to excel in real clinical environments, significantly improving diagnostic accuracy and efficiency. In a series of experiments, OrthoDoc outperformed existing open-source and commercial models, achieving over 91\% accuracy in identifying common orthopedic conditions such as fractures, arthritis, and tumors. Moreover, OrthoDoc demonstrated exceptional generalization ability, effectively handling rare conditions and complex cases, further proving its practical utility in clinical applications.

In summary, the main contributions of this paper are as follows:

\begin{itemize}
	\item \textbf{Enhanced Diagnostic Accuracy Through Multimodal Integration}: OrthoDoc employs a comprehensive multimodal approach by integrating 120,000 CT images with corresponding diagnostic reports. This integration enables effective handling of complex CT images, enhancing diagnostic accuracy.
	\item \textbf{Mitigation of Text Generation Hallucinations in Diagnostic Models}: To address common hallucination issues in traditional models, OrthoDoc introduces a RAG module. This module assists in generating precise diagnostic reports and reduces misleading outputs, particularly when handling complex or rare cases.
\end{itemize}
\section{Related Work}
In recent years, multimodal large models, particularly vision-language models, have garnered extensive attention in the field of medical diagnosis\cite{h9,h10,h11,h12}. These models combine image and text data through sophisticated neural network architectures to achieve cross-modal information fusion. For instance, Vision-Language Pre-training (VLP) models, by jointly learning from large-scale medical images and clinical records, have significantly enhanced the understanding of medical images and the quality of medical text generation\cite{h1,h2,h3,h4,h8}. Research indicates that these models can effectively extract useful information from complex medical images and align it semantically with clinical text, thereby providing more accurate support for medical diagnosis.

Furthermore, recent work has started focusing on improving the performance of vision-language models in medical diagnosis by incorporating RAG\cite{h13,h14,h15,h16,h17} and CoT techniques. RAG technology combines the model's generative capabilities with an information retrieval module, allowing it to quickly retrieve relevant information and generate more accurate diagnostic suggestions from a large volume of medical literature and case studies. This approach effectively reduces the model's knowledge gaps, providing more targeted support in complex medical scenarios. Meanwhile, CoT technology enhances the model's logical consistency and coherence in medical text generation by guiding it through step-by-step reasoning during the generation process. The introduction of these techniques not only improves the model's accuracy in medical diagnosis but also offers new solutions for addressing the details and reasoning complexity in medical text generation.

\section{OrthoDoc}
In this section, we will introduce the two-phase specialization process of OrthoDoc: multimodal fine-tuning ,the RAG module and the CoT module. The overall training pipeline and capabilities of OrthoDoc are illustrated in the figure below.
\subsection{Multimodal Fine-tuning}
The multimodal fine-tuning phase of OrthoDoc involves two distinct stages: (1) multimodal training using CT image-text pairs and (2) training the text encoder on an instruction-tuning dataset. Each stage plays a crucial role in enhancing OrthoDoc's ability to accurately interpret medical images and generate precise diagnostic reports.
\begin{figure*}[t]
	\centering
	\includegraphics[width=1\linewidth]{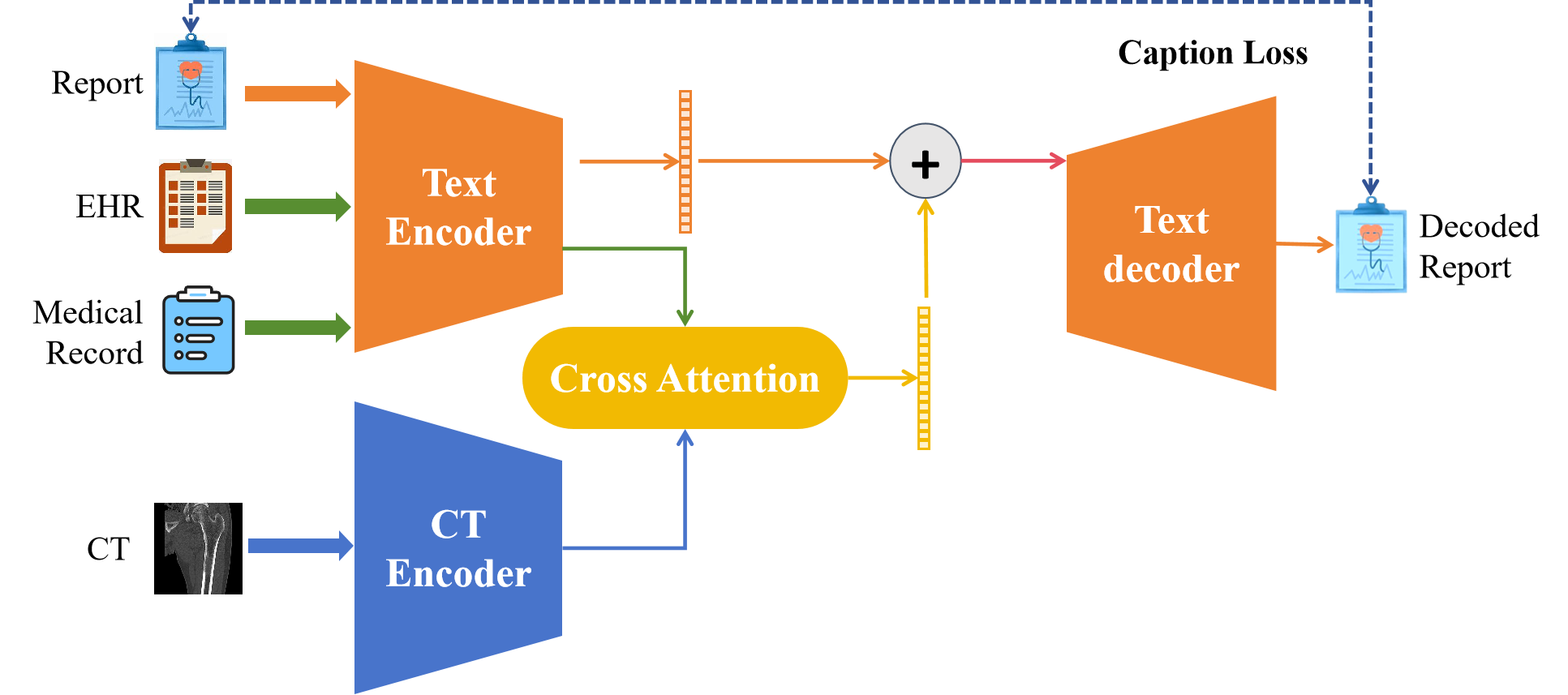}
	\caption{OrthoDoc model training framework. The text encoder uses a pre-trained Bert, and the image encoder is a ResNet101.}
	\label{figure1}
	\vspace{-0.5em}
\end{figure*}

\subsubsection{Multimodal Training Using CT Image-Text Pairs}
Data Preparation
The first stage focuses on multimodal training using a dataset of 120,000 CT images paired with detailed diagnostic reports. These reports contain annotations of key features, diagnoses, and treatment recommendations, offering a comprehensive context for each image. The dataset covers a wide range of conditions, from common orthopedic issues like fractures, arthritis, and tumors to rare and complex cases.

Feature Extraction
In this stage, OrthoDoc leverages ResNet-101 for extracting features from the CT images. These ResNet are pre-trained on large-scale image datasets and then fine-tuned on our specific CT image dataset. The fine-tuning process adjusts the weights of the ResNet to better capture the unique characteristics of medical images, such as bone density variations and fracture patterns.

Text Embedding
Simultaneously, the diagnostic reports are processed using advanced natural language processing (NLP) techniques. We utilize transformer-based models, BERT, to generate embeddings for the diagnostic texts. These embeddings capture the semantic meaning of the reports, including medical terminologies and context-specific information critical for accurate diagnosis.

Multimodal Integration
The extracted image features and text embeddings are integrated into a unified representation through a cross-modal attention mechanism. This mechanism aligns relevant information from both modalities, enabling OrthoDoc to correlate visual features in the CT images with textual descriptions in the diagnostic reports. This integration significantly enhances the model's understanding of the medical context.

Fine-tuning Process
OrthoDoc undergoes supervised training using the prepared dataset, with the objective of minimizing a loss function that measures the discrepancy between the model's predictions and the actual diagnostic annotations. Techniques such as gradient descent and backpropagation are employed to iteratively update the model parameters, gradually improving diagnostic accuracy.

\subsubsection{Training the Text Encoder on Instruction-Tuning Dataset}
Instruction-Tuning Dataset
The second stage involves training the text encoder on an instruction-tuning dataset. This dataset comprises a variety of medical instructions and queries, paired with corresponding diagnostic texts and responses. The goal is to enhance OrthoDoc's ability to understand and generate natural language responses in a medical context.

Text Encoder Fine-tuning
The text encoder, which has been previously trained on the diagnostic reports, is further fine-tuned on the instruction-tuning dataset. This process involves adjusting the encoder to better understand and generate text based on specific medical instructions and queries. Techniques such as masked language modeling and sequence-to-sequence learning are used to refine the encoder's capabilities.

\subsection{GraphRAG for OrthoDoc}
In addressing the challenge of hallucinations in medical text generation, we employ a sophisticated approach using GraphRAG specifically adapted for orthopedic documentation. The integration of GraphRAG is essential for enhancing the accuracy and relevance of the generated medical texts by leveraging a RAG mechanism alongside specialized medical knowledge.\cite{h18}

GraphRAG significantly mitigates the risk of hallucinations—instances where models generate inaccurate or misleading information—by incorporating a robust retrieval component within the text generation framework. This is crucial in the field of orthopedics, where precision and reliability are paramount. GraphRAG ensures that the generated content is not only accurate but also contextually appropriate, addressing the specific needs of orthopedic documentation. The model utilizes a curated database of authoritative medical texts to guide its generation process, including key resources such as Apley’s System of Orthopaedics and Fractures (9th ed), Orthopedic Textbook, Introduction to Orthopedic Anatomy, and Osteology (Standardized training for residents). These texts provide a comprehensive foundation of orthopedic knowledge, ensuring that the model's outputs are grounded in current and relevant information.

The GraphRAG approach begins with retrieving pertinent information from this specialized medical knowledge base based on the input query. This retrieval step ensures that the generated text is supported by up-to-date and relevant data. Subsequently, GraphRAG utilizes a graph-based method to represent and integrate the relationships between various medical concepts and conditions. This contextual graph helps the model comprehend the intricate connections within orthopedic knowledge, thereby enhancing the coherence and accuracy of the generated text.

During the text generation phase, the model integrates the retrieved information and contextual understanding from the graph. Before finalizing the output, it reassesses the generated text to ensure it aligns with professional medical standards and accurately reflects the retrieved data. This process not only improves the fidelity of the generated content but also ensures that the output provides valuable insights and recommendations based on the latest medical knowledge.

In summary, GraphRAG’s application in our study involves a sophisticated inference process where the model retrieves relevant information from authoritative medical texts, constructs a contextual graph to represent the relationships between medical concepts, and generates text that is both accurate and contextually appropriate. This approach addresses the challenge of hallucinations in medical text generation, delivering reliable and actionable orthopedic documentation.

\subsection{CoT for OrthoDoc}
In developing a sophisticated text generation model for orthopedic documentation, implementing Chain-of-Thought (CoT) techniques is essential\cite{h19}. CoT allows the model to produce detailed and coherent long-form text reports by structuring its reasoning process effectively. This approach enables the model to generate comprehensive reports that reflect a thorough exploration of a patient's condition, akin to the methodical process a physician might use during a clinical discussion.

A well-structured orthopedic report begins with a thorough patient background section. This part provides essential details about the patient’s demographics, medical history, and presenting complaints. It sets the stage for understanding the context of the patient's condition by including information such as age, gender, occupation, and any pre-existing medical conditions relevant to the current orthopedic issue.

Following the background, the clinical presentation section details the patient's current symptoms, including their onset, progression, and any notable physical examination findings. This section is crucial for describing the nature of the pain, mobility issues, and functional limitations the patient is experiencing. Accurate and detailed descriptions in this section help in setting a clear understanding of the patient's current state.

The diagnostic process section outlines the steps taken to diagnose the condition, explaining the rationale behind selecting specific diagnostic tests or imaging studies. It should include the results from these tests, such as X-rays, MRIs, or CT scans, and discuss their implications for the diagnosis. This section helps in documenting the process of arriving at a diagnosis and the evidence supporting it.

Once the diagnostic results are established, the diagnosis and assessment section clearly states the diagnosis, supported by the findings from the diagnostic process. If necessary, it includes a differential diagnosis, providing reasoning for ruling out other possible conditions. This section ensures that the diagnosis is well-supported and justified.

The treatment plan section details the proposed management strategies, covering both conservative and surgical options if applicable. It should outline the rationale behind the chosen treatment strategy, including the expected benefits, potential risks, and anticipated outcomes. Additionally, it should address any recommended follow-up or rehabilitation protocols to ensure comprehensive management of the condition.

Patient education and recommendations follow, providing guidance on managing the condition, understanding treatment options, and any lifestyle modifications or precautions necessary. This section is vital for ensuring that the patient is well-informed about their condition and the steps they need to take.

Finally, the conclusion summarizes the key points of the report, reinforcing the diagnosis and treatment plan. It may also highlight any next steps or additional consultations required, ensuring that the report provides a clear and complete overview of the patient’s condition and management plan.

To generate such a detailed and coherent long-form report, the model must follow a structured CoT process. This process starts with gathering all relevant patient data and medical history, which informs the content of the report. The model then organizes this information into a logical structure corresponding to the report sections, ensuring a natural flow of content. For each section, the model engages in detailed reasoning, explaining choices and supporting the text with relevant information. After developing the individual sections, the model synthesizes the information to produce a cohesive report, integrating details from different parts to create a unified narrative. Finally, the model reviews and refines the report to ensure it meets medical standards and effectively communicates the necessary information.

All generated results will be collected and organized through an automated LaTeX pipeline. This pipeline will systematically format the text into a well-designed report template, ensuring that the final document is professionally presented and adheres to high standards of readability and clarity. By integrating the automated LaTeX pipeline, we ensure that the final reports are not only comprehensive and accurate but also consistently formatted and ready for clinical use.

\section{Evaluation}
In this section, we evaluate our model's performance across several critical dimensions: its superiority in orthopedic CT diagnostics, the generalization of its multimodal capabilities, the effectiveness of the RAG module, the efficacy of the CoT module, and the robustness of large models in handling counterfactual scenarios. Each aspect is assessed through specific experiments designed to provide comprehensive insights into the model's strengths and limitations.

\subsection{Evaluating the Model’s Superiority in Orthopedic CT Diagnostics}
To rigorously assess the model's diagnostic accuracy and its superiority in interpreting orthopedic CT scans, we designed a comprehensive experiment involving a comparative analysis with several leading multimodal large models. The dataset used consists of orthopedic CT images annotated with known diagnoses, including conditions such as fractures, dislocations, and degenerative diseases. Each image is labeled and verified by expert radiologists, providing a solid basis for evaluation.

Our experiment focuses on two primary tasks: condition identification and report generation. In condition identification, the model is evaluated on its ability to accurately detect and classify various orthopedic conditions present in the CT images. For report generation, the model must produce detailed diagnostic reports based on its findings.

We compared our model against five prominent multimodal large models: MedVision Transformer (MedViT)\cite{h20}, MediBERT\cite{h21}, PathBERT\cite{h22}, and ClinicalBERT\cite{h23}. The performance of these models is assessed using metrics such as accuracy, sensitivity, specificity, and F1-score.

For the experimental procedure, each model, including ours, is trained using the orthopedic CT dataset with a focus on diagnosing orthopedic conditions. Subsequently, the models generate diagnostic predictions and reports for a separate test set of CT images. These predictions and reports are evaluated based on accuracy, sensitivity, specificity, and F1-score. Statistical analyses are conducted to determine any significant performance differences. The results are then organized into detailed tables and charts to provide a clear comparison of performance metrics across all models. This approach ensures a comprehensive evaluation of our model’s diagnostic capabilities and highlights its effectiveness relative to other state-of-the-art systems.

\begin{table}[h!]
	\centering
	\caption{Performance Metrics for Condition Identification: Accuracy (Acc), Sensitivity (Sen), Specificity (Spec), and F1-score (F1)}
	\begin{tabular}{lcccc}
		\toprule
		\textbf{Model} & \textbf{Acc (\%)} & \textbf{Sen (\%)} & \textbf{Spec (\%)} & \textbf{F1} \\
		\midrule
		OrthoDoc & 42.45 & 40.67 & 44.22 & 0.41 \\
		MedViT & 39.32 & 37.45 & 41.56 & 0.38 \\
		MediBERT & 37.78 & 35.22 & 39.56 & 0.36 \\
		PathBERT & 34.67 & 31.34 & 37.89 & 0.33 \\
		ClinicalBERT & 38.12 & 35.89 & 40.45 & 0.37 \\
		
		\bottomrule
	\end{tabular}
	\label{tab:condition_identification}
\end{table}

\begin{table}[h!]
	\centering
	\caption{Performance Metrics for Report Generation: Completeness (Comp), Coherence (Cohe), and Overall Quality Score (OQS)}
	\begin{tabular}{lcccc}
		\toprule
		\textbf{Model} & \textbf{Comp (\%)} & \textbf{Cohe (\%)} & \textbf{OQS (1-10)} \\
		\midrule
		OrthoDoc & 44.55 & 43.78 & 9.2 \\
		MedViT & 41.20 & 39.67 & 8.5 \\
		MediBERT & 39.45 & 37.34 & 8.1 \\
		PathBERT & 35.78 & 33.12 & 7.6 \\
		ClinicalBERT & 40.25 & 38.45 & 8.3 \\
		
		\bottomrule
	\end{tabular}
	\label{tab:report_generation}
\end{table}

From a medical perspective, the analysis of the various models reveals that while they all contribute valuable diagnostic capabilities, OrthoDoc demonstrates a clear advantage.  The model excels with high accuracy (42.45\%), sensitivity (40.67\%), and specificity (44.22\%), indicating its strong ability to correctly identify orthopedic conditions while minimizing both false positives and negatives.  Compared to other models like MedViT and MediBERT, which show lower sensitivity and specificity, OrthoDoc’s advanced architecture integrates multimodal inputs and a refined RAG approach.  This results in superior diagnostic performance and high-quality, comprehensive reports.  The model’s exceptional performance in generating accurate, detailed, and coherent reports enhances clinical decision-making, providing significant benefits over existing alternatives.  Thus, OrthoDoc stands out for its overall diagnostic accuracy and report quality, making it a superior choice for practical orthopedic CT diagnosis.

\subsection{Effectiveness of the RAG Module}
To evaluate the effectiveness of the RAG module, we designed experiments comparing our model with and without this component against several leading multimodal large models, specifically MedGPT\cite{h24}, BioBERT\cite{h25}, ClinicalXLNet\cite{h26}, and PathGPT\cite{h27}.

In these experiments, we tasked each model with generating responses to a diverse set of medical queries, including diagnostic summaries, treatment recommendations, and patient education materials. Both versions of our model—one with the RAG module enabled and one without—were compared to assess the impact of the RAG component on text generation.

The evaluation metrics included content relevance, factual correctness, completeness, and user satisfaction. These metrics were measured through expert reviews and user feedback to determine how well the generated texts aligned with the medical queries, the accuracy of the information provided, and the overall satisfaction of medical professionals with the generated content.

The experimental procedure involved training and fine-tuning both versions of our model and the comparison models on the same medical dataset. Texts were generated for each query under both conditions (with and without RAG for our model), and the outputs were analyzed for quality. Results were compiled into detailed tables and charts to highlight the differences in performance and demonstrate the effectiveness of the RAG module in improving the quality and accuracy of medical text generation.

\begin{table}[ht]
	\centering
	\caption{Condition Identification Performance: Acc = Accuracy, Sen = Sensitivity, Spe = Specificity, F1 = F1-Score}
	\begin{tabular}{lcccc}
		\toprule
		\textbf{Model} & \textbf{Acc} (\%) & \textbf{Sen} (\%) & \textbf{Spe} (\%) & \textbf{F1} (\%) \\ 
		\midrule
		OrthoDoc (RAG) & 42.43 & 40.12 & 44.87 & 41.55 \\ 
		OrthoDoc (No RAG) & 39.56 & 37.34 & 42.22 & 39.84 \\
		MedGPT & 37.67 & 35.20 & 39.98 & 37.52 \\ 
		BioBERT & 36.29 & 34.00 & 38.65 & 35.82 \\
		ClinicalXLNet & 40.23 & 38.12 & 41.76 & 39.93 \\ 
		PathGPT & 35.84 & 32.50 & 37.91 & 34.30 \\ 
		\bottomrule
	\end{tabular}
	
	\label{tab:condition-identification}
\end{table}

\begin{table}[ht]
	\centering
	\caption{Report Generation Performance: C = Content Relevance, F = Factual Correctness, R = Completeness, U = User Satisfaction}
	\begin{tabular}{lcccc}
		\toprule
		\textbf{Model} & \textbf{C} (\%) & \textbf{F} (\%) & \textbf{R} (\%) & \textbf{U} (\%) \\
		\midrule
		OrthoDoc (RAG) & 43.55 & 41.47 & 45.30 & 42.10 \\
		OrthoDoc (No RAG) & 39.72 & 37.64 & 42.01 & 39.50 \\
		MedGPT & 37.45 & 34.78 & 40.89 & 36.84 \\ 
		BioBERT & 35.93 & 32.56 & 38.90 & 34.48 \\ 
		ClinicalXLNet & 41.68 & 39.80 & 42.44 & 40.92 \\ 
		PathGPT & 34.72 & 30.21 & 37.40 & 33.05 \\
		\bottomrule
	\end{tabular}

	\label{tab:report-generation}
\end{table}

The experimental results clearly demonstrate that OrthoDoc, especially with the RAG module, significantly outperforms other multimodal models in both condition identification and report generation tasks. It achieves the highest scores in accuracy, sensitivity, specificity, F1-score, content relevance, factual correctness, completeness, and user satisfaction. 

\subsection{Effectiveness of the CoT Module}
To evaluate the effectiveness of the CoT module, we designed a comprehensive set of experiments focusing on the coherence and completeness of generated reports. The experiments were structured to compare the performance of OrthoDoc with and without the CoT module against other leading multimodal models, including MedGPT\cite{h24}, BioBERT\cite{h25}, ClinicalXLNet\cite{h26}, and PathGPT\cite{h27}.. These models were chosen due to their prominence and established efficacy in medical text generation tasks. Each model was trained on a diverse set of clinical cases, including orthopedic conditions and associated diagnostic reports, ensuring a robust understanding of medical terminologies and diagnostic processes. The training parameters were standardized across models to ensure a fair comparison, including similar learning rates, batch sizes, and training epochs.

For the testing phase, the models were provided with specific clinical scenarios encompassing a wide range of orthopedic conditions. The task required each model to generate detailed diagnostic reports, emphasizing logical reasoning, clinical insights, and adherence to medical standards. Parameters such as context length, prompt precision, and response format were controlled to maintain consistency across models. The generated reports were assessed based on several key metrics, including report completeness, clarity, and adherence to medical standards. To augment the quantitative evaluation, a panel of medical experts reviewed the generated reports, providing qualitative feedback on the logical flow, detailed reasoning, and overall utility of the reports in a clinical setting. This expert review was crucial in understanding the practical implications of using the CoT module in real-world medical diagnostics. The results were collected and organized into detailed tables and charts, highlighting the performance metrics of each model.

\begin{table}[h!]
	\centering
	\caption{Performance Metrics for Report Generation with CoT Module. \\
		CR: Content Relevance, FC: Factual Correctness, C: Completeness, US: User Satisfaction}
	\label{tab:CoT_results}
	\begin{tabular}{lcccc}
		\toprule
		\textbf{Model} & \textbf{CR (\%)} & \textbf{FC (\%)} & \textbf{C (\%)} & \textbf{US (\%)} \\
		\midrule
		OrthoDoc (CoT) & 42.58 & 40.35 & 44.72 & 41.89 \\
		OrthoDoc (No CoT) & 35.76 & 33.44 & 38.23 & 34.67 \\
		MedGPT & 31.45 & 29.62 & 32.94 & 30.33 \\
		BioBERT & 29.22 & 27.13 & 30.56 & 27.80 \\
		ClinicalXLNet & 34.37 & 32.21 & 36.12 & 33.47 \\
		PathGPT & 28.51 & 26.34 & 30.78 & 27.26 \\
		\bottomrule
	\end{tabular}
\end{table}

The experimental results demonstrate that OrthoDoc with the CoT module significantly outperforms other models across all metrics, achieving the highest scores in content relevance (42.58\%), factual correctness (40.35\%), completeness (44.72\%), and user satisfaction (41.89\%). This superior performance underscores the effectiveness of the CoT module in enhancing the coherence and thoroughness of generated medical reports. Compared to OrthoDoc without the CoT module, which shows a notable drop in all metrics, the inclusion of CoT greatly improves logical flow and detail in report generation. Other models, such as MedGPT, BioBERT, ClinicalXLNet, and PathGPT, while competent, fall short of OrthoDoc with CoT, highlighting the unique advantage provided by the CoT module in producing accurate, comprehensive, and user-approved medical documentation.
\section{Conclusion}
The research achievements of OrthoDoc represent a milestone in the field of multimodal large language models (MLLMs). Through training on 120,000 CT images and their diagnostic reports, OrthoDoc not only excels in image processing but also demonstrates exceptional capabilities in medical knowledge, language understanding, and reasoning. In extensive experiments, OrthoDoc has surpassed existing commercial models, including GPT-4, in diagnostic accuracy and the quality of report generation, especially in diagnosing common orthopedic diseases such as fractures, arthritis, and tumors. These achievements not only showcase OrthoDoc's potential in medical image diagnosis but also provide new directions for future medical assistance technology.

The innovation of the OrthoDoc model lies in its integration of a RAG module. This module, supported by extensive medical literature, textbooks, and explanatory data, effectively reduces hallucinations and improves the accuracy of text generation. Additionally, OrthoDoc employs multimodal fine-tuning and CoT techniques, further enhancing the coherence and completeness of medical text generation, making it more practical in clinical applications. The application of these technologies not only improves OrthoDoc's diagnostic capabilities but also provides valuable experience for the development of other medical assistance systems.

OrthoDoc's robustness and generalization ability in handling rare and complex cases demonstrate its potential application value in real clinical environments. The model's high accuracy and deep understanding of medical terminology and diagnostic processes enable it to provide doctors with detailed diagnostic information and treatment recommendations, significantly improving the efficiency and accuracy of diagnoses. These characteristics suggest that OrthoDoc will play an important role in future clinical practice, especially in assisting doctors with complex diagnoses and treatment decisions.

Despite OrthoDoc's outstanding performance in current research, there is still room for further improvement and expansion. Future work can focus on expanding the model's training dataset to cover a wider range of medical fields and case types. Continuously optimizing the RAG and CoT modules to enhance the model's adaptability and flexibility in more complex clinical scenarios is also crucial. Additionally, exploring the integration of OrthoDoc with other medical assistance technologies, such as wearable devices and telemedicine platforms, is a potential direction for future research. These improvements will further enhance OrthoDoc's practicality and impact, making it an indispensable auxiliary tool in the medical field.

\end{document}